\documentclass[journal, 10pt]{IEEEtran}
\usepackage[figuresright]{rotating}
\usepackage{amssymb}
\usepackage{lipsum}
\usepackage{graphicx}
\usepackage{caption}
\usepackage{subcaption}
\ifCLASSINFOpdf
\else
\fi
\usepackage[cmex10]{amsmath}
\usepackage{array}
\usepackage{graphicx}
\usepackage{caption}
\usepackage{subcaption}
\usepackage[lofdepth,lotdepth]{subfig}
\usepackage{mdwmath}
\usepackage{mdwtab}
\usepackage{graphicx}
\usepackage{algpseudocode}
\usepackage{algorithm}
\ifCLASSINFOpdf
 
\else
  
\fi

\begin{document}
%
\title{Information Estimation with Node Placement Strategy in 3D Wireless Sensor Networks} 

\author{ Jyotirmoy Karjee\IEEEauthorrefmark{1} and  H.S Jamadagni$\dag$
	 \thanks{\IEEEauthorrefmark{1} Corresponding author}
	\thanks{Embedded Systems and Robotics Group, TCS Research and Innovation, Bangalore, INDIA; $\dag$Department of Electronic Systems Engineering, Indian Institute of Science, Bangalore, INDIA}  \thanks{Email: jyotirmoy.karjee@tcs.com, hsjam@dese.iisc.ernet.in}}

\maketitle

\begin{abstract}

The cluster formation in Three Dimensional Wireless Sensor Networks (3D-WSN) give rise to overlapping of signals due to spherical sensing range which leads to information redundancy in the network. To address this problem, we develop a sensing algorithm for 3D-WSN based on dodecahedron topology which we call Three Dimensional Distributed Clustering (3D-DC) algorithm. Using 3D-DC algorithm in 3D-WSN, accurate information extraction appears to be a major challenge due to the environmental noise where a Cluster Head (CH) node gathers and estimates information in each dodecahedron cluster. Hence, to extract precise information in each dodecahedron cluster, we propose Three Dimensional Information Estimation (3D-IE) algorithm. Moreover, Node deployment strategy also plays an important factor to maximize information accuracy in 3D-WSN. In most cases, sensor nodes are deployed deterministically or randomly. But both the deployment scenario are not aware of where to exactly place the sensor nodes to extract more information in terms of \emph{accuracy}. Therefore, placing nodes in its appropriate positions in 3D-WSN is a challenging task. We propose a Three Dimensional Node Placement (3D-NP) algorithm which can find the possible nodes and their deployment strategy to maximize information accuracy in the network. We perform simulations using MATLAB to validate the 3D-DC, 3D-IE and 3D-NP, algorithms respectively. \\


\end{abstract}

\section{Introduction}

Wireless Sensor Networks (WSN) \cite{jk1} have made sufficiently great attention in the field of signal processing and wireless communications. Sensor nodes are made up of MEMS \cite{jk2}, \cite{jk3} and smaller in size having low processing speed, generally used for sensing and extracting data in a network. WSN are used in military applications, under water \cite{jk4}, environmental monitoring \cite{jk5}, health care applications and many more. Sensor nodes are capable to measure an event features (like temperature, pressure, humidity, etc.). In WSN, a group of sensor nodes perform similar sensing task of an event features (like measuring moisture content of agricultural field, reading temperature of an indoor room, measuring seismic event, detecting event target, detecting fire in a forest etc). An event is defined as a physical occurrence which is measured by groups of sensor nodes in a network. Sensor nodes continuously measure the physical phenomenon of an event and report measured information to the sink node.

 Sensor nodes does collaborative sensing task in two dimensional \cite{jk6}-\cite{jk11} networks, where nodes are deployed in agricultural field, forest, etc. However, WSN can also be applied  in three dimensional \cite{jk15}-\cite{jk17} space by measuring movement and behavior of birds or insects and fixing sensor nodes with probes in underwater \cite{jk4}. The information collected by sensor nodes deployed in three dimensional space are generally spatially correlated within a network. Sensor node uses correlated information to form clusters.  In a 3D clusters based sensing model \cite{jk15}, the sensing range of a cluster is considered as sphere. Since the sensing range of clusters are spherical, it creates tessellate (overlapping or gapping) of sensing coverage as shown in Figure 1. In this figure, three clusters are given namely $A$, $B$, $C$ respectively. Each cluster have its own neighboring nodes within its spherical sensing range. As shown in figure, cluster A and B respectively overlaps in its spherical range which leads to increase the bandwidth utilization in 3D network. Moreover, in between cluster's $A$, $B$ and $C$ respectively, there is a gap (tessellate) among spherical sensing range which leads to poor sensing coverage in 3D networks. 

\begin{figure}
\centering
\includegraphics[width=0.30\textwidth]{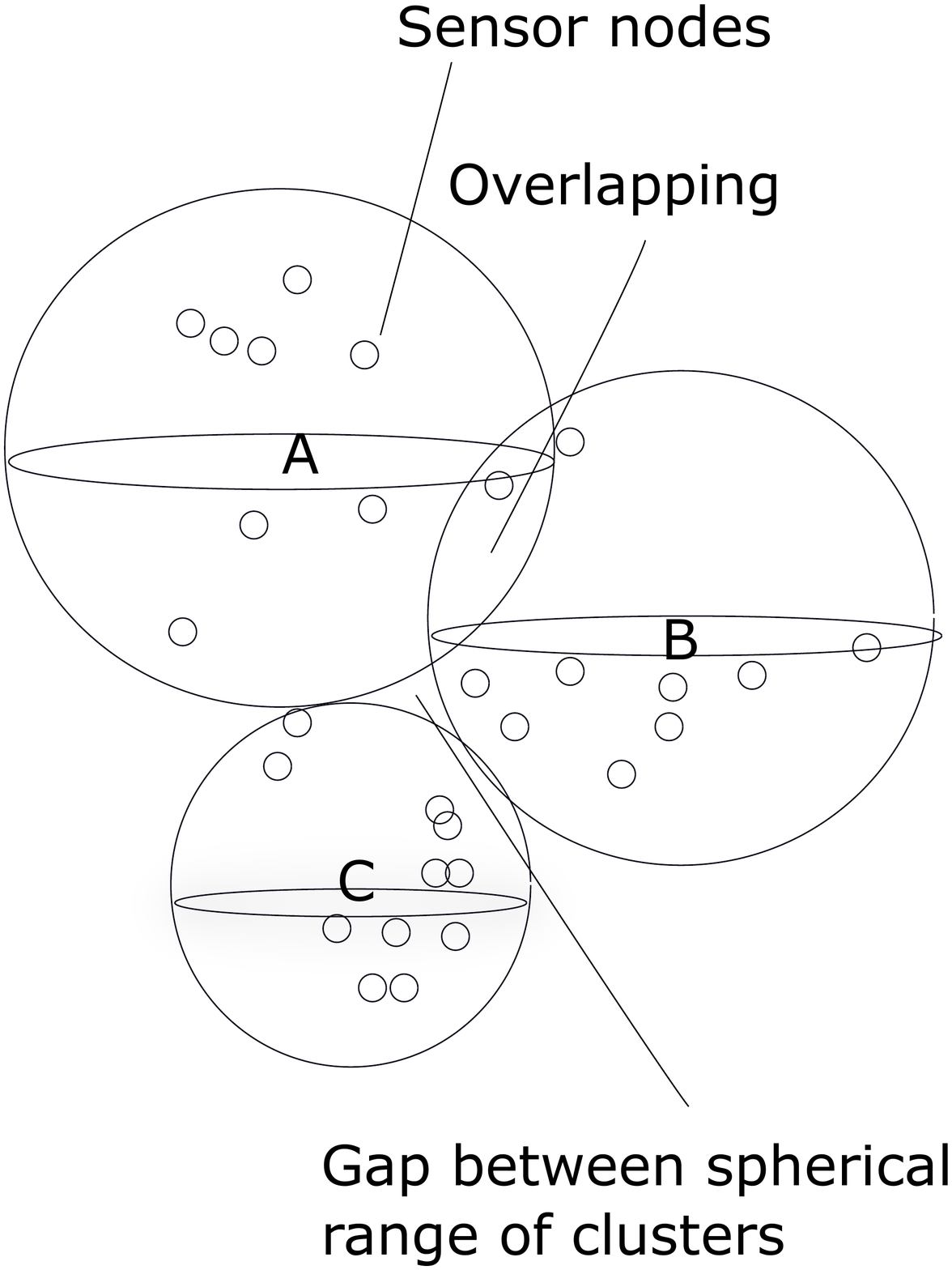}
\caption{Tessellate among spherical range of clusters }
\end{figure}

 The assessment of information is a major issue in 3D-WSN as sensor nodes have a limitation of having low processing power and have sort communication range. Based upon the statistical knowledge of data \cite{jk14}, sensor node extract data (with prior knowledge or without prior knowledge of data statistics) from the environment, process it and finally transmits to the Base Station (BS). While extracting data, in some cases sensor networks assume that they have prior knowledge of data statistics (like variance and covariance of sensed data) in the network. But most often, sensor nodes extract data in a regular interval of time without having prior knowledge of data statistics. Thus, sensor nodes extract statistical information from the environment under two scenarios. Firstly, sensor nodes extract information within a network under priori knowledge of statistical information (e.g variance and covariance) of the environment. For example, sensor networks have prior knowledge of variance and covariance of temperature of an indoor room.  In WSN, extracting information without having prior knowledge of information statistics is a challenging task in three dimensional space. Hence, in this paper, we focus on extracting information in 3D space without having prior knowledge of signal statistics. Here, sensor nodes observe and sense information dynamically and transmit it to CH node or base station.

Real time sensor data is monitored continuously over temporal and spatial \cite{jk6} domains, respectively. The information extracted by sensor nodes are generally spatially correlated within a network. As the node density increases, the spatially proximal \cite{jk12}, \cite{jk13} observation of correlated information among the nodes also increases in 3D-WSN. Since, spatial information received at the sink node are highly correlated, it increases information redundancy in the network. Hence, it is required to  minimized information redundancy in WSNs. Moreover, in each cluster, cluster Head (CH) node calculates a distortion factor \cite{jk6}-\cite{jk11} (information accuracy) in 3D space. Distortion factor or information accuracy \cite{jk8} is  defined as the degree of closeness of measured data to its actual sensed data in a network. Data accuracy \cite{jk6}-\cite{jk11} models are developed under spatial data correlation. These models calculates a minimum set of sensor nodes which are sufficient to give the desired data accuracy level as achieved by the whole network. Data accuracy can be model under online \cite{jk14} data extraction to select optimal sensor nodes in the network. Generally, information accuracy is maximized to get accurate information but leads to increase in energy consumption in the network. To balance these, a trade off \cite{jk10} between information accuracy and energy consumption is built to increase the lifetime of sensor networks. It also  selects an optimal sensor nodes, thereby reducing the communication overhead in the network. 

 In Figure 2, a 3D space is considered having three coordinates where we are interested to measure the information profile of each sensor node. In this 3D space lets assuming ten sensor nodes are deployed randomly. Out of ten sensor nodes, lets say six sensor nodes have reached the desired maximum information level. Hence node number: $1$, $3$, $4$, $7$, $8$ and $10$ are considered in the network to maximize the information accuracy. The sensor nodes with bellow of that desired information level may have lower information profile. In reality, it may be possible that node id: $1$, $3$, $4$, $7$, $8$ and $10$ are directly put under the sunlight, due to which the signal variance goes above the desired information accuracy level, where as node id: $2$, $5$, $6$ and $9$ are placed under the shade of a tree. This may be the reason due to which the signal variance of the sensor nodes are bellow that desired information accuracy level of that specific location. Hence, it may be recommended that nodes with lower information profile in that specific location are added with extra deployed sensor nodes so that a desirable information accuracy can be extracted on that specific location.

\begin{figure}
\centering
\includegraphics[width=0.30\textwidth]{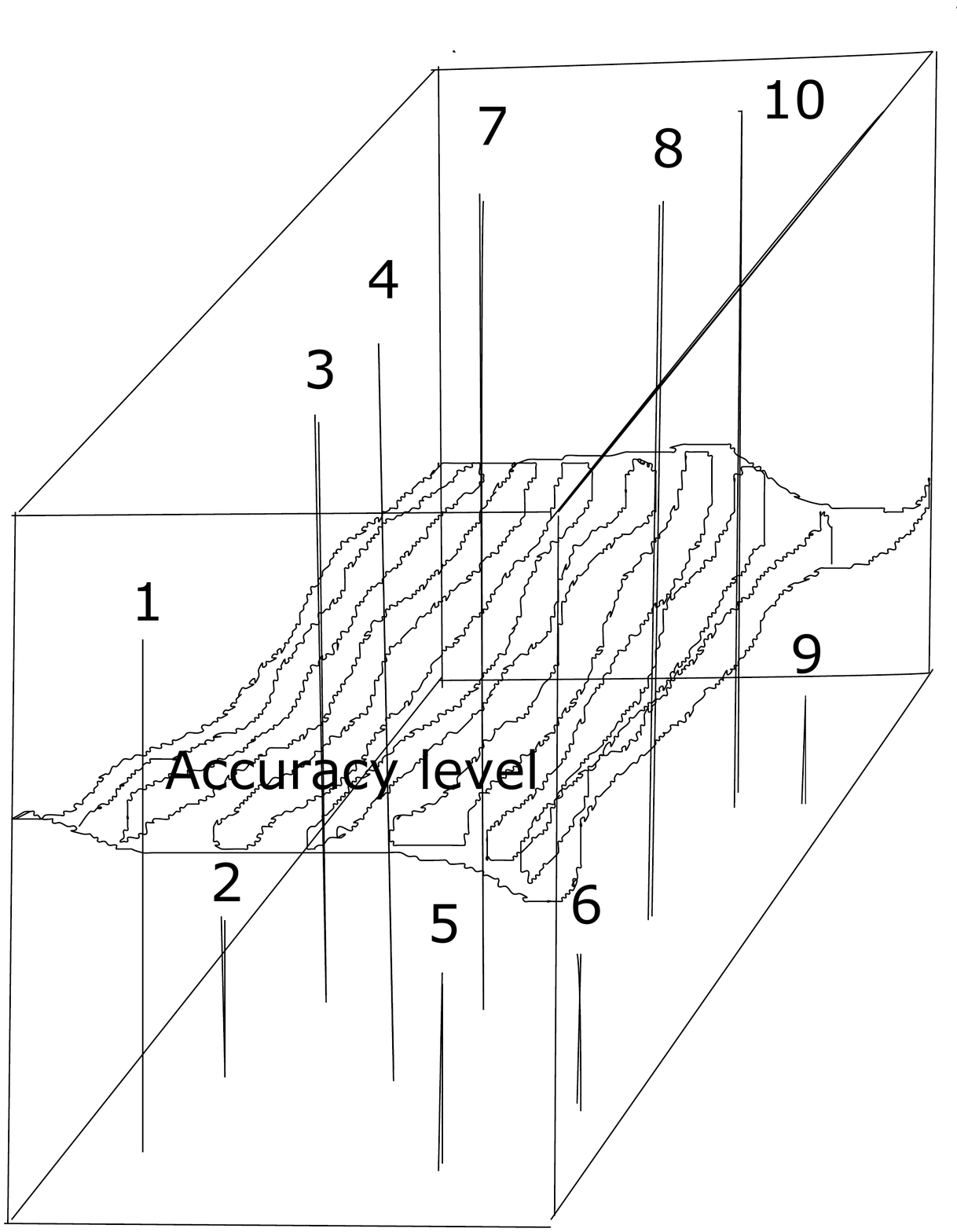}
\caption{Temperature profile extracted by nodes in 3D space}
\end{figure}

To extract accurate information, nodes deployment in 3D-space plays a crucial role in wireless sensor networks. Sensor nodes can be deployed in two ways: (a) It can be deterministically deployed by manually putting it in the sensing field by human effort. But these type of nodes deployment is inefficient in terms of time and cost. (b) It can be randomly deployed in the sensing field by aerial or human effort. But both the procedure, we are not aware of extracting maximum information in a specific location due to exact placement of sensor nodes in 3D wireless sensor networks. Therefore, placement of sensor nodes plays a vital role in terms of maximizing the information accuracy with node placement strategy in wireless sensor networks. In this paper, we consider the application of environmental monitoring to measure temperature data in 3D-WSN. The paper is organized as follows: In section II, we discuss the related works. In section III, we have addressed the problem definitions of our work. In section IV, we describe the mathematical models of 3D-DC, 3D-IE and 3D-NP algorithms respectively. Validations and simulations of the propose work are presented in Section V. Finally, we conclude our work in Section VI.

\section{Related Work}

In \cite{jkA27}, author's proposed the confident information coverage model where sensor nodes are deployed such that they cover the entire geographical area in wireless sensor networks. In \cite{jkA28}, author's proposed an optimal node placement patterns based on confident information coverage (CIC) considering the collaboration of sensor information and spatial correlation of physical data. In \cite{jkA29}, author's investigate optimal node placement for long belt coverage in wireless sensor networks. Moreover, in \cite{jkA30}-\cite{jkA32}, authors discussed the information coverage problem implemented in two dimensional wireless sensor networks, but none of the above works illustrate the information coverage problem in three dimensional networks, which is still a open challenge in communication networks. The above literatures, also haven't addressed the information coverage problem, in three dimensional distributed scenario. However, many clustering algorithms \cite{jk10}-\cite{jk13} are proposed for WSN in two dimensional space. But none of the work addresses the formation of clusters using jointly sensing nodes in 3D space. In \cite{jk15}, a 3D spherical based sensing model is proposed, but the model creates tessellate (overlapping or gapping) in between the spherical ranges which utilizes more bandwidth in the network. However, in \cite{jk16} authors addressed the problem to model the sensing coverage by tessellating polyhedra but lags the formation of cluster in 3D-WSN. Moreover, various data estimation models \cite{jk6}-\cite{jk11}, \cite{jk15}-\cite{jk17}, are developed with \emph{a-prior} knowledge of data statistics in two dimensional space and three dimensional space, but none of the works focus on estimating information without having \emph{a-priori} knowledge of information statistics in 3D-WSN. In \cite{jk6}-\cite{jk11}, data accuracy models are developed under spatial data correlation. These models calculates a minimum set of sensor nodes which are sufficient to give the desired data accuracy level as achieved by the whole network. Similarly, in \cite{jk14}, \cite{jk1411} data accuracy model is developed under online data extraction to select optimal sensor nodes in the network. In \cite{jk10}, a trade off between data accuracy and energy consumption is developed to select an optimal number of sensor nodes, thereby reducing the communication overhead in the network. But none of the works have determine which sensor nodes are to be selected for achieving maximum information accuracy in the network. In the above literature, sensor nodes are either deterministically or randomly deployed in the sensing field to find an optimal set of sensor nodes maintaining a desired information accuracy level. Therefore, It also doesn't explore which sensor nodes are to be selected. Moreover, it doesn't clarifies where exactly is to be place sensor nodes in 3D-space to get maximum information accuracy.

\section{Problem Formulations with Specific Use-case}

In this section, we emphasize to formulate our problem definitions as follows.

\emph{1. Formation of distributed clusters using dodecahedran topology for better sensing coverage is a subject of interest in 3D-WSN}: In 3D-WSN, sensing coverage of clusters are generally modelled as spherical topology. But spherical range of clusters form tessellate (overlapping or gaping) space within clusters. This lead to poor sensing coverage (due to space among spherical range of cluster) or more bandwidth utilization (due to overlapping of spherical range of clusters) in 3D-WSN. Moreover deployment of more nodes in the sensing range creates unnecessary data redundancy in 3D-WSN due to spatially correlated data. To overcome this problem, we develop Three Dimensional Distributed Clustering (3D-DC) algorithm using dodecahedron topology as shown in Figure 3. In this figure, three clusters namely $D$, $E$ and $F$ are shown with dodecahedron topology. Each dodecahedron clusters has its own neighboring sensor nodes within its range. As shown in figure, there may not have tessellate between the clusters which reduce the information redundancy in 3D networks.

\begin{figure}
\centering
\includegraphics[width=0.30\textwidth]{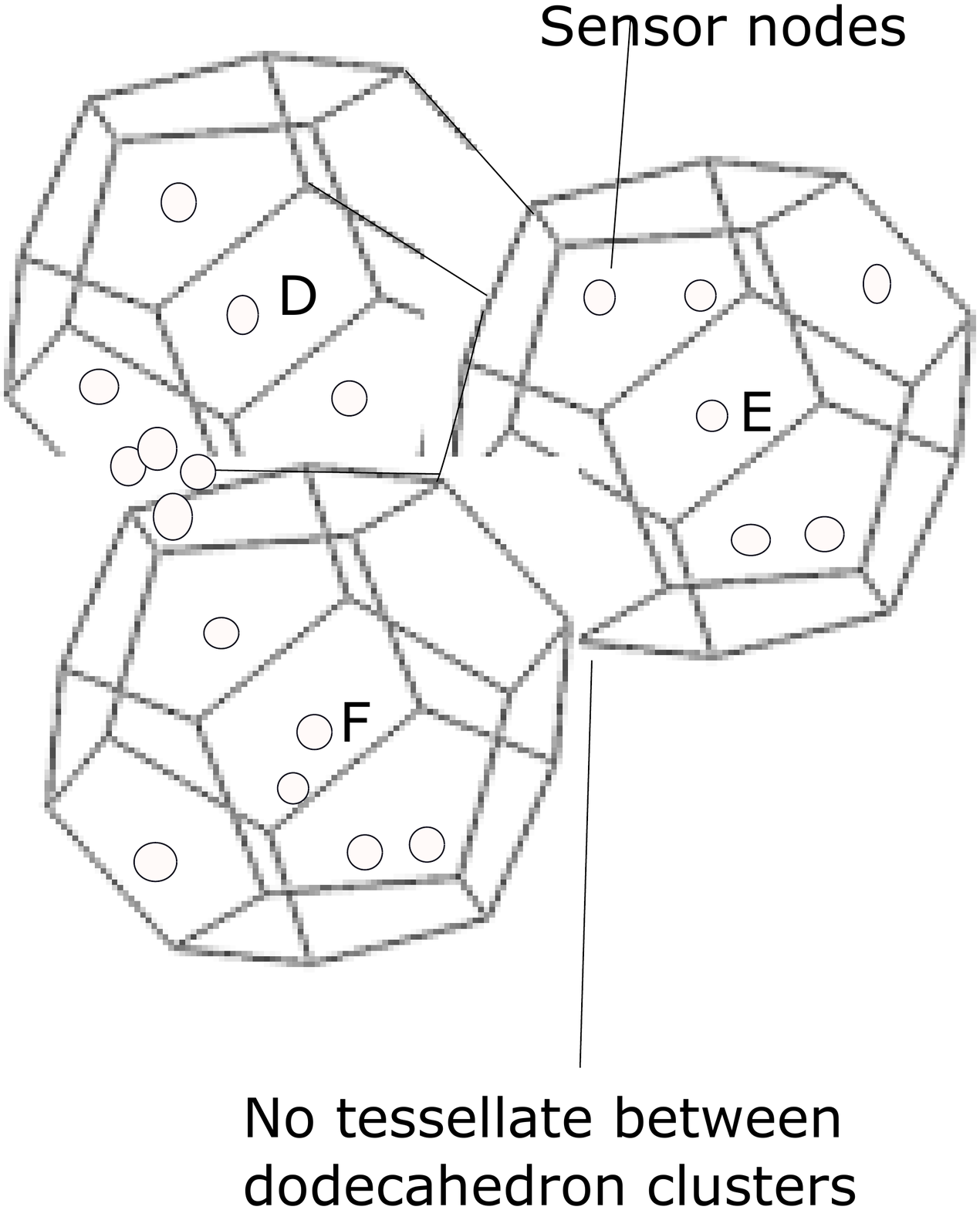}
\caption{Dodecahedron range of clusters having no tessellate }
\end{figure}

\emph{2. Information estimation without having priori knowledge of signal statistics (i.e online information estimation) is a challenging task in 3D-WSN: } 
Sensor nodes form distributed clusters in two dimensional space and three dimensional space using spatially correlated data among sensor nodes. Data correlation among sensor nodes have both priori and without priori information statistics (i.e known and unknown variance or covariance) in the network. Using the knowledge of information statistics, information estimation models are developed in two dimensional WSN. Moreover, information estimation with prior knowledge of signal statistics is implemented in 3D-WSN but non of the work emphasize on estimating information without priori knowledge of signal statistics in 3D-WSN. Hence, information estimation in 3D-WSN is a challenging task (e.g estimating accurate information in under-water).

 \emph{3. To extract maximum information in a sensor network, maximum nodes deployment is necessary for some specific sensing region where sensor signal penetration may be difficult}:
As discussed in the previous section, sensor nodes can be deployed deterministically and randomly in a sensing field to extract information in a network. But both the deployment scenarios fails to extract accurate information in a network as we are not aware where to place sensor nodes to get more information. To clarify this, lets assume sensor nodes deployment scenario in a forest to sense temperature in a specific sensing region like 3D-space. Throughout the paper, we assume deployment of sensor nodes in a 3D-space to measure temperature as an example.
As nodes are deployed randomly, it may be possible that at some sensing region (3D-space), placing more sensor nodes are required to maximize the information. Generally, sensor signal penetration is less in dense forest (tree dominated) region in 3D space. In such region, more number of sensor nodes can be deployed to extract more information. But it is unclear as what exactly the upper bound of sensor nodes to be deployed in such a region to extract adequate information. Here, optimal node placement plays an important role to maximize information accuracy where sensor signal penetration is difficult.

\emph{4. Sensor nodes are placed closer to the target event to maximize the information accuracy }:
We take another scenario, where we deploy sensor nodes closer to the event \cite{jk6} target to maximize information accuracy in 3D network. Placing sensor nodes closer to the target event may extract more information than placing sensor node far apart from the event. But it doesn't give any clear idea at what minimum distance, nodes are to be deployed with respect to the target event and by which topology nodes can be deployed to extract more information in a network. It may be possible that placing sensor nodes closer to target event may cause more noise distortion \cite{jk8} due to signal interference. In this case, a deployment topology in 3D space can be developed by which sensor nodes can be placed with minimum distance with respect to the target event to extract information without noise distortion.

 \emph{5. We give emphasize on temporal data collection as there may be variation of data collected depending upon the node placement in 3D space}:
Generally, data collected in a sensor networks have spatio-temporal correlation among them. Some times, it is more important to give emphasise on temporal information than spatial information for node deployment strategy in 3D space. Considering a network scenario in a forest, where a set of sensor nodes are deployed under the shade of dense trees and another set of nodes are placed directly under sunlight. Assuming that both the set of nodes are placed close to each other in the sensing region. Since both set of sensor nodes are closer and collects an event feature (i.e temperature), there is a difference in variation of temporal data collected by them as one set of nodes is placed under shade of tree and other placed directly under sunlight. In this case, we deploy sensor node such that more nodes are to be placed under shade of a tree due to week sensor signal where as less number of sensor nodes are placed directly under sunlight. Still, it is unclear that what is the maximum number (upper bound) of nodes to be placed under shade of a tree and minimum number (lower bound) of nodes to be deployed under sunlight to get information accuracy in a network. Thus optimal deployment of nodes is necessary for this scenario.

\emph{6. Consideration may be given to spatial data correlation among sensor nodes to determine minimum number of sensor nodes required to maximize information accuracy}:
We consider a network scenario, where a set of sensor nodes are close to each other and directly collects data from sunlight in a sensing region. In this case, sensor nodes have almost similar signal variance and therefore, consideration is given to spatial data correlation among sensor nodes in a network. We make use of spatial data correlation among sensor nodes to determine optimal sensor nodes in a network with maximum information accuracy.

\section{Mathematical Model}

In this section, we develop the following mathematical basis of our models: firstly, we develop a three dimensional distributed clustering (3D-DC) algorithm in 3D space. Secondly, with in each cluster, we develop three dimensional information estimation (3D-IE) model at CH node of the cluster. 3D-DC and 3D-IE algorithms are the extension of work presented in \cite{jk15}.  Finally, three dimentional node placement (3D-NP) algorithm is developed for the network which maintains a desired information accuracy.

\subsection{Cluster Formation in Three Dimensional Space}

Considering a static source event which propagates its signal spherically in 3D space \cite{jk15}. We ignore the attenuation or propagation delay of signal from the source event. Assuming that the nodes which are within the spherical signal range can only sense the source event. The information extracted by sensor nodes for the source event can be modelled as spatial data correlation using exponential model \cite{jk18}, \cite{jk19} given by  $C_{exp}(d)= exp^{\left(-\frac{|d|^{\alpha}}{\theta} \right)} $ where $\alpha =1$, $\alpha$ is the smoothness parameter which gives the geometric attributes of the propagated signals, $\theta$ is the range parameter which determines the delay of signal w.r.t to distance and $d$ is the distance among the source event to the sensor nodes in 3D space. Using exponential model, we develop the 3D-DC algorithm as follows: Data correlation between source event to number of sensor nodes in 3D space; Data correlation model among sensor nodes; Formation of distributed clusters in 3D space. The 3D-DC is explained in Algorithm 1.

\begin{algorithm}
\caption{3D Distributed Clustering Algorithm}

\textbf{Require:} A source event $S$ occurred and sensor nodes $i$ and $j$ are randomly deployed in a three dimensional space.

\textbf{Returns:} Formation of clusters in three dimensional space.

\begin{algorithmic}[1]

\State Start

\State Compute information correlation among source event $\emph{s}$ and sensor node \emph{i}

 \State  Define a threshold $\tau_{e}$ to verify whether the information is strongly correlated among $s$ and $s_i$ or weakly correlated

\State  If the information correlation is strong,
then calculate euclidean distance $d_{s,i}$ among event $s$ source to $i$th sensor nodes

\State   Compare euclidian distance $d_{s,i}$, with the radius  of the spherical range of event source $r_e$

\State   Compute volume of spherical range of event source $ V_e$ from step 4

\State Assuming the sensing range of sensor nodes \emph{i} and \emph{j} as dodecahedron which lie within $\emph{V}_{e}$.

\State Calculate radius of circumscribed sphere of a node  as $ r_n$

\State Calculate the volume of dodecahedron sensing range of a node as $V_n$.

\State Find spatial information correlation among $s_i$ and  $s_j$.

 \State Define a threshold value $\tau_n$ such that information is strongly correlated.

 \State  Calculate circumscribed spherical radius of a node  as $r_n $ where vertices's of dodecahedron touches the spherical range.

  \State Update $V_n$ using step 11, where $V_n \subseteq V_e$

\State  To form clusters in 3D space, assuming set of sensor node $\textit{A}$ with dodecahedron sensing range lie within $V_e$.

\State Within sensing range $V_e$, $\forall i \in \textit{A}, \hspace{0.1in}  \textup{let}  \hspace{0.1in}  \mathcal{E}(i)=\{j \in \textit{A} : d(i,j) \leq r_n, i\neq j  \} $ where \emph{d(i,j}) is the euclidian distance between \emph{i} and \emph{j} having dodecahedron sensing range.

\State  $\mathcal{G} = \{j \in \mathcal{L} : \mathcal{E}(j)=\max\mathcal{E}(i)\}, i \in \mathcal{L} $ , we define
$d_{max}(i)=\displaystyle \max_{j\in \mathcal{E}(i)}d(i,j)$, where \emph{d(i,j)} is the Euclidian distance between \emph{i} and \emph{j} within $v_n$

\State  Consider $\mathcal{P}=\displaystyle \arg \min_{i\in \mathcal{G}}d_{max}(i) $ and $C=C\cup \{(\mathcal{P},\mathcal{E}(\mathcal{P}) )\}$ where
$C$ is the set of cluster of dodecahedron range 

\State  For each {$\emph{c}$=($\emph{x}_c$ , $\emph{y}_c$)},verifies the Euclidian distance between \emph{x} to $s$ to select  $C=\displaystyle \min_{i\in \mathcal{G}} d_{x_s,S}$ with $\emph{c} \in C$

\State  $\textit{A}=\textit{A}-\{\mathcal{P}\}-\mathcal{E}\{\mathcal{P}\} $.

\State  If $\textit{A} \neq \{ \phi \} $, go to step 16.

\State End
\end{algorithmic}
\end{algorithm}

\emph{(a). Data correlation between source event to number of sensor nodes in 3D space}: 
We are assuming that a source event $s$ occur in 3D space. The event $s$ propagates its signal spherically in 3D space. The sensor node $i$ which are situated within the effect of spherical range $s$, can form information correlation among $s$ and node $i$ such that $\emph{s}$ and \emph{i}: $\rho[\emph{s},\emph{s}_i]=$ $C_{exp}(d_{s,i})=  exp^{\left(-\frac{|d_{s,s_i}|^{\alpha}}{\theta} \right)}$ for $\alpha=1$. We define a threshold value $\tau_e$ \cite{jk9,jk10} to determine the strength of information correlation.  If $\rho[\emph{s},\emph{s}_i] \geq \tau_e $, information shows strong correlation among $s$ and $s_i$ otherwise weakly correlated i.e $C_{exp}(d_{s,i})= exp^{\left(-\frac{|d|^{\alpha}}{\theta} \right)} \geq \tau_e $. If it shows strong information correlation, then the euclidian distance among event source to $i$th sensor nodes is given as $d_{s,i} \leq  \left( \theta \hspace{0.04in} log \left(\frac{1}{\tau_e} \right) \right)^{1/\alpha}$ in 3D space. We compare euclidian distance $d_{s,i}\leq  \left( \theta \hspace{0.04in} log \left(\frac{1}{\tau_e} \right) \right)^{1/\alpha}$, with spherical source event radius $r_e$ is given by $\left(\frac{3V_e } {4\pi} \right)^{1/3}$ to get $\left(\frac{3V_e } {4\pi} \right)^{1/3}=\left( \theta \hspace{0.04in} log \left(\frac{1}{\tau_e} \right) \right)^{1/\alpha}$. Finally, we compute volume of spherical range of event source as $ V_e=\left( \frac{4 \pi } {3 }\right)\left(\theta  \hspace{0.04in} log \left(\frac{1}{\tau_e} \right) \right)^{3/\alpha} $.
 The parameters $\tau_e$, $\theta$ and $\alpha$ are dependent on $v_e$. If $\tau_e$ increases, $V_e$ decreases exponentially \cite{jk15} with fixed value of $\theta$ and $\alpha$.

\emph{(b). Data correlation model among sensor nodes:} In second phase of 3D-DC algorithm, we illustrate spatial information correlation among sensor nodes which belongs to the spherical range of source event $s$. Thus, sensing nodes which occurs with in volume $V_e$ can only form jointly sensing nodes to explore spatial correlation between them. We assume that sensing range of nodes are dodecahedron \cite{jk16} in shape which are within the range $V_e$. The sensing range of a node with dodecahedron shape is circumscribed in a spherical range where the vertices's of dodecahedron touches the spherical range. The radius of circumscribed sphere of a node is given as $ r_n=\frac{\nu \sqrt{3}}{4}(1+ \sqrt{5})$ where $\nu$ is the edge range of dodecahedron. The volume of dodecahedron sensing range of a node is given as $
V_n=\frac{\nu}{4}(15 + 7 \sqrt{5})$. Considering the value of $r_n$ in $V_n$, we the volume of dodecahedron sensing range $V_n=\frac{4}{\sqrt{3}} \frac{(15 + 7 \sqrt{5})}{(1+ \sqrt{5})}r_n $. We are interested to find spatial information correlation among dodecahedron sensing range of sensor nodes $s_i$ and  $s_j$ which is given as $\rho[\emph{s}_i,\emph{s}_j]=C_{(.)}(d_{s_i,s_j})$. We define another threshold value $\tau_n$ to determine spatial data correlation among nodes $s_i$ and $s_j$ with dodecahedron sensing range.
 If $\rho[\emph{s}_i,\emph{s}_j] \geq \tau_n $, information are strongly correlated else weakly correlated within $V_e$. For strongly correlated information i.e $\rho[\emph{s}_i,\emph{s}_j] \geq \tau_n $, we get $d_{s_i,s_j} \leq  \left( \theta \hspace{0.04in} log \left(\frac{1}{\tau_n} \right) \right)^{1/\alpha} \leq d_{s,s_i}$.
 Comparing this with three dimensional euclidian distance between nodes $i$ and $j$, we get the circumscribed spherical radius of a node as $r_n \leq  \left( \theta \hspace{0.04in} log \left(\frac{1}{\tau_n} \right) \right)^{1/\alpha}$ where the vertices of dodecahedron touches the spherical range. This means $r_n=\frac{\nu \sqrt{3}}{4}(1+ \sqrt{5})=\left( \theta \hspace{0.04in} log \left(\frac{1}{\tau_n} \right) \right)^{1/\alpha}$. Therefore $V_n=\frac{4}{ \sqrt{3}} \frac{(15 + 7 \sqrt{5})}{(1+ \sqrt{5})}\left( \theta \hspace{0.04in} log \left(\frac{1}{\tau_n} \right) \right)^{1/\alpha} $ where $V_n \subseteq V_e$. The size of $V_n$ depends upon the value of $\tau_n$, $\theta$ and $\alpha$ and always $V_n \leq V_e$. For a fixed value of $\theta$ and $\alpha$, if we increase $\tau_n$, the size $V_n$ decreases exponentially thereby the average number of distributed clusters decreases exponentially within $V_e$ \cite{jk15}.

\emph{(c). Formation of distributed clusters in 3D space}:
 In the third phase of 3D-DC algorithm, the sensing nodes co-operate each other to form clusters. In \cite{jk15}, authors proposed 3D-ESCC algorithm where jointly sensing nodes form clusters having spherical sensing range. The major drawback of 3D-ESCC algorithm is that the spherical range of clusters form tessellate (overlapping or gapping) among different clusters. Hence it is essential to develop a sensing coverage model which can fulfill the gap among two clusters and remove the overlapping of sensing coverage among clusters. To overcome these problems, we propose a sensing model in this paper where sensing coverage of clusters have dodecahedron \footnote{Theoretically sensor nodes with uniform dodecahedron sensing range can remove the tessellate space for cluster formation, where as in algorithm 1, we consider that there are tessellate among clusters since they are not uniform in 3D space.  } topology  which may overcome the tessellate space among the clusters. For the formation of clusters, each sensor node finds its dodecahedron sensing range according to Algorithm 1 within the source event volume $V_e$. The dodecahedron sensing range of a sensor node having maximum number of neighbouring nodes within its range can form the first cluster. Similarly, the formation of next clusters is considered to have decreasing number of sensor nodes within its dodecahedron sensing range and so on. For example, if a sensor node $i$ form the first cluster having five neighboring sensor nodes within its dodecahedron sensing range then the next sensor node $j$ can form cluster having less than five neighbouring sensor nodes within its dodecahedron sensing range without tessellate in 3D-space. Moreover, if sensor nodes $i$ and $j$ have same number of neighbouring nodes within its dodecahedron sensing ranges, then which form the first cluster is a subject of interest? The question is trivial. In this case, sensor nodes $i$ and $j$ finds the farthest position of neighboring nodes from the centre position of nodes $i$ and $j$ with its dodecahedron sensing range respectively, then it calculates which of the neighboring node is closer to its position $i$ and $j$ nodes respectively. The neighbouring node with minimum distance from the centre position of nodes $i$ and $j$, can form the first cluster, since closer the sensing node, the information correlation among sensing node increases, the detailed formation of clustering algorithm is explained in Algorithm 1 (step 14 to step 20).

\subsection{Data Estimation in Three Dimensional Space }

Initially, we deploy sensor nodes in three dimensional (3D) space. In 3D space, sensor nodes form distributed cluster according to 3D-DC algorithm within $V_e$. Here, we define Three Dimensional Information Estimation (3D-IE) algorithm by adopting 3D-ESCC \cite{jk15} algorithm and extending the work when CH node of cluster extract information in dynamic condition (without having prior knowledge of information). In each cluster, each sensor node transmits its observed data to its respective CH node of the cluster. In 3D space, each cluster is responsible for estimating single source point event $s$. Let $u_c^i$ is the observation done by a cluster $C$ where $i$ denotes the number of cluster formed in 3D space. Hence, observation done by all the clusters is given by $\sum^{p}_{i=1} u_c^i$ where $p$ is the total number of clusters in 3D space.

We consider a single cluster $C$, where each sensor node $j$ can sensed the data $s_j(t)$ from the source point event $s$ at a time interval $t$ under additive white gaussian noise (AWGN) given by

\begin{equation}\label{eq:jkx}
u_{c_j}(t)=s_j(t)e^{kwt}+\alpha_j(t)  \hspace{0.2in}  \textup{\emph{j=1,2,. . , m}}    \hspace{0.2in}
\end{equation}

where \emph{k}=$\sqrt{-1}$, carrier frequency $\omega=2\pi f$ and $m=$ number of sensing nodes in a cluster $C$. We denote $\Delta t$ as the total time required for signal prorogation from point event source $s$ to each sensor node \emph{j} of cluster $C$ with distance $\emph{d}$ given by

\begin{equation}\label{eq:jkx}
\Delta t=\frac{d_{S,S_{c_j}} }{\Omega}=\frac{2\pi d_{S,S_{c_j}}}{\omega_c \lambda}
\end{equation}

where $\Omega$ is the velocity of data prorogation in atmospheric medium and $\lambda$ is the wavelength of data signal \cite{jk15}.

\textbf{Data Propagation:} For wireless sensor networks, we assume two channels by which sensor data can propagate: sensing channel and communication channel. Since sensor nodes are deployed in three dimensional space, each sensor node can extract data through sensing channel. Once the sensor node extract raw data, they transmits data among the networks through the communication channel. Thus, communication channel is any medium (e.g air, water, space) by which sensor nodes can transmit data among them in a cluster $C$.

If each sensor node $j$ observed data as $v_{c_j}(t)=s_j(t)e^{kwt} + \alpha_j (t)$ at time interval $t$, then the observation done by another node $l$ with delay varying time $\Delta t$  is given as
 $v_{c_l}(t)=s_l(t+\Delta t)e^{k\omega (t + \Delta t}+ \alpha_l(t)$ in any medium. According to slow varying \cite{jk15}, \cite{jk20} data signal, we consider $s(t+\Delta t)\cong s(t)$. Hence, in a cluster $C$, each node observed data as

 \begin{equation}\label{eq:jkx}
 v_{c_l}(t)=s_l(t)e^{k\omega t}e^{k \frac{2\pi d_{S,S_{c_l}}}{ \lambda} }
 \end{equation}

 We conclude that data prorogation from node $j$ with observation $u_{c_j}$ is delayed with $\Delta t$ with observation $v_{c_l}$ within a cluster $C$. In a cluster $C$, the data received by each sensor nodes are converted to baseband. This makes the carrier signal $e^{-kwt}$ to be vanished.
Considering the $qth$ count of observations done by sensor node at time $t$ is given as

\begin{equation}
u_{c_l}(t)=s_l(t)e^{k \frac{2 \pi q}{\lambda} d_{S,S_{c_l}} }  +  \alpha_j(t)  \hspace{0.2in}  \emph{q=0,1,2,. . .,m-1}
\end{equation}

We describe a linear model using Gauss-Markov Theorem \cite{jk20}, at each CH node of cluster given by

\begin{equation}
U(t)=L.s(t)+ \Psi(t)
\end{equation}

where

$L=\left(1  \hspace{0.1in}  e^{j \frac{2 \pi }{\lambda} d_{S,S_1} }   \hspace{0.1in}  e^{j \frac{4 \pi }{\lambda} d_{S,S_2 }}  \hspace{0.1in}    \cdot   \hspace{0.1in}      \cdot \hspace{0.1in} e^{j \frac{2 \pi (m-1)}{\lambda} d_{S,S_{m-1}} } \right)^T$,

$\Psi(t)=\left(  \hspace{0.05in} \psi_0(t)  \hspace{0.1in}    \psi _1(t) \hspace{0.1in}  \psi_2(t) \hspace{0.1in}  \cdot  \hspace{0.1in} \cdot  \hspace{0.1in} \psi_{m-1}(t)  \hspace{0.05in}  \right)^T$

 We define $L$ as a $m \times 1$ vector of sensor nodes locations, $s(t)$ is the actual source event to be estimated at time $t$, $\Psi(t)$ in a additive with gaussian noise at time $t$ with zero mean and covariance $\Lambda$. We are interested to find how much accurate information, we can extract at CH node of a cluster under noisy environment. Thus, we formulate an estimator using mean square error \cite{jk20} represented as

\begin{equation}
I(m)=E[s(t)-\hat{s}(t)]^2
\end{equation}

We construct $\hat{s}(t)$ using best linear unbiased estimator (BLUE) as $\hat{s}(t)=\frac{L^T \Lambda^{-1}U(t)}{L^T\Lambda^{-1}L}$. We assume that sensor nodes extract reliable information under additive white gaussian noise which is uncorrelated with $s$ with zero mean and form covariance matrix $\Lambda=\sigma^2_{{n}_i}I$. Finally, $m$ sensor nodes in cluster are extracting accurate information at CH node under noisy correlation given as

\begin{equation}
\hat{s}(t)=\frac{1}{m } \sum^{m-1}_{n=0} \sum^{m}_{i=1}    U_{n}(t)e^{-j \frac{2 \pi q}{\lambda} d_{S,S_i} }
\end{equation}

To get the normalized information at the CH node of a cluster, we modify

\begin{equation}
I_{A}(m)=1-\frac{I(m)}{E[s^2(t)]}=\frac{1}{E[s^2(t)]}[2E[s(t)\hat{s}(t)]-E[\hat{s}^2(t) ]
\end{equation}

To get the normalized information at CH node of a cluster, we modify \cite{jk4} as;
$Cov[s,s_i]=\sigma^2_{s}Corr[s,s_i]=\sigma^2_{s}\rho(s,s_i)=\sigma^2_{s}C_{exp}(d_{s,s_i})$. Similarly,
$Cov[s_i,s_j]=\sigma^2_{s}Corr[s_i,s_j]=\sigma^2_{s}\rho(s_i,s_j)=\sigma^2_{s}C_{exp}(d_{s_i,s_j})$. The $d_{s,s_i}$ is defined as $ d_{s,s_i}=\parallel s - s_j \parallel$ as euclidian distance among source $s$ to number of node $i$ in a cluster. Similarly, $d_{s_i,s_j}$ is given as $ d_{s_i,s_j}=\parallel s_i - s_j \parallel$ as euclidian distance among nodes $i$ and $j$ at time interval time $t$. $\sigma^2_{s_i}$ is the signal variance extended by sensor nodes $i$ and $j$ at time interval $t$. Finally, we calculate the information accuracy using 3D-IE in a CH node of cluster using known signal statistics given as

 $I_{A}({m})={\frac{1}{m}}{\left(2{\sum_{i=1}^{m}\rho_{S,S_i}}\right)}-{\frac{1}{m^2}} \left(
   {\sum_{i=1}^{m}}{\sum_{j\neq i}^{m}}{\rho_{S_i,S_j}} \right)$

 \begin{equation}
   - {\frac{1 }{ m^2} }
   \left(\frac{m\sigma_S^2 + \sum_{i=1}^{m}\sigma_{n_i}^2} {\sigma_{S }^2} \right)
\end{equation}

\emph{Case study when nodes fails to operate in 3D networks:} In 3D space, it may be possible that any of the nodes in network fails to operate due to environmental conditions \cite{jk15}. In such situation, information accuracy may degrade. To overcome this problem, data prediction can be done for that specific node failure in the 3D space. Assuming there are \emph{O} sensor nodes within sensing range $V_e$ which maximizes the information accuracy with locations at $L_{1}$, $L_{2}$, . . $L_{O}$. The sensor observation done by nodes can be expressed as
 $\emph{s}(\emph{L}_{1})$, $\emph{s}(\emph{L}_{2})$, . . $\emph{s}(\emph{L}_{m})$. Let say with \emph{O} sensor nodes, \emph{n} sensor nodes are dead and \emph{m} are active for doing communication process such that \emph{O=n+m}. Our goal is to predict observed data for \emph{n} dead nodes. The unobserved value of dead nodes can be represented as  ${s_d}(\emph{L}_{1})$, ${s_d}(\emph{L}_{2})$, . . ${s_d}(\emph{L}_{n})$ at locations $L_{1}$, $L_{2}$, . . $L_{n}$ where the coordinates of sensor positions are known. Hence, in a 3D space, the sensor observation considered to predict information $\int \emph{s}(\emph{L})\emph{dL}$ represented as

\begin{equation}\label{eq:11}
s(L)=\chi +I(L)  
\end{equation}

where $\chi$ is an unknown factor and \emph{I}(\emph{L}) is intrinsically stationary, IS \cite{jk22}. Taking into consideration (\ref{eq:11})
for specific location $Z_{0}$ in 3D space, our aim is to find unobserved information values ${s_d}(\emph{Z}_{1}), {s_d}(\emph{Z}_{1}), ....., {s_d}(\emph{Z}_{n}) $.  To evaluate the best predicted of unobserved information of ${s}_d(L_{1})$, ${s}_d(L_{2})$, . . ${S}_d(L_{n})$ from the following observed information ${s}(L_{n+1}), {s}(L_{n+2}), ......s(L_m)$ in 3D space given by

\begin{equation}\label{eq:jk2}
\hat{s}(L_d)=E[s(L_d)|s_m]
\end{equation}

where $L_d=L_1, L_2,........L_n$ and $s_m=[s(L_{n+1}), s(L_{n+2}), . . \emph{s}(L_{m})]$. The prediction $\hat{\emph{s}}(L_d)$ is defined as the average of the observed information extracted by \emph{ O} sensor nodes in 3D space as

\begin{equation}\label{eq:jk2}
\hat{s}(L_{d})=\frac{1}{O}\sum^{O}_{i=1}\emph{S}(\emph{L}_{i}) \hspace{0.2in} i=1,2,. . . O
\end{equation}

where $n+m=O$. Using minimum minimum mean square, we design the predictor as

$\hspace{0.6in} P_d=E[s(L_{d})-\hat{s}(L_{d})]^2 \hspace{0.2in}$

To get better estimation of predicted information, we normalized the information as

$ P_{d(Nor)}=1-\frac{P_d}{E[s^2(L_{d})]}$

\begin{equation}
\hspace{0.7in} =\frac{2}{O}\sum^{O}_{i=1}\rho(d_{L_d,L_i})-\frac{1}{m^2} \sum^{O}_{i=1} \sum^{O}_{j\neq i}\rho(d_{L_i,L_j)}
\end{equation}

$P_{d(Nor)}$ is the normalized form of predictor which predicts information for sensor node within the sensing range $V_e$.

 \subsection{Node Placement in 3D Networks}

 In this subsection, we are interested to find which nodes are to be selected to maximize information estimation within the sensing range $V_e$. Hence, we propose three dimensional Node Placement (3D-NP) algorithm, which can find an optimal nodes using global optimization problem \cite{jk23}- \cite{jk26} in 3D space. Using 3D-NP algorithm each node position is search in the geographical location which actually maximizes information accuracy in the network. Hence, the goal is to maximize information in that specific location to be searched where a set of sensor nodes collaboratively does the operation. In some cases, sensor nodes placed in 3D network doesn't have prior knowledge of location where the information is maximized and which sensor nodes are to be selected for maximizing the information. It may be possible that sensor nodes have prior knowledge about its locations where the information can be maximized. For example, sensor nodes placed directly under sunlight gains more information accuracy than the nodes placed under a dense forest area. Hence, the effective strategy is to find the sensor nodes which are placed in such a location where the information accuracy may be more (i.e nodes placed directly under sunlight). In this optimization problem, each sensor node have a searching space. Hence, in 3D space  each node have a cost value which are calculated using a cost function which is to be optimized. For this optimization problem, sensor nodes does the searching operation by updating its iterations. In each iteration, each node signal variance is updated by the following two \emph{best} signal variances. The first variance of cost value we called $\sigma^2_b$. Another signal variance is tracked using the optimization problem obtained by any sensor nodes within the whole 3D network. The best signal variance is called the global variance ($\sigma^2_{gb}$). The sensor nodes participated in the whole 3D network within its neighbouring nodes, we assume the best variance is the local signal variance ($\sigma^2_l$) in the network. After finding the two best signal variances, sensor node updates its signal variance to maximize information accuracy in its location \cite{jk23} as following.

\begin{equation}\label{eq:jkk1}
I_A=I_A + \phi_1 (\sigma^2_b - \sigma^2_p) + \phi_2 (\sigma^2_{gb} - \sigma^2_p)
\end{equation}

\begin{equation}\label{eq:jkk2}
\sigma^2_p=\sigma^2_p + I_A
\end{equation}

where $I_A$ is the information accuracy of sensor nodes, $\sigma^2_p$ is the present signal variance of sensor node and $\phi_1$, $\phi_2$ are the adaptation factor to learn signal variance w.r.t to time. 3D-NP algorithm is summarize as follows in Algorithm 2.

\begin{algorithm}
\caption{Three Dimentional Node Placement Algorithm }

\begin{algorithmic}[1]

\State For each sensor node
\State Initialize signal variance of sensor node in 3D space
\State End
\State  For each sensor node
\State  Calculate cost function value which is better than the best signal variance ($\sigma^2_b$)
\State  Set present cost value as the new best signal variance ($\sigma^2_b$)
\State End
\State Select sensor node with the best cost function value (signal variance) among all sensor nodes in 3D network as the global variance ($\sigma^2_{gb}$)
\State For each sensor node
\State Calculate the maximum information accuracy using (\ref{eq:jkk1})
\State  Update the position of sensor nodes using (\ref{eq:jkk2}) in 3D space
\State End

\end{algorithmic}
\end{algorithm}

Thus a set of sensor nodes can be selected with in a 3D network which maximizes information accuracy using 3D-NP algorithm.

\section{Simulations and Validations}

In this section, we validate 3D-DC, 3D-IE and 3D-NP algorithms respectively by taking sensor data from Intel Berkeley Research Lab \cite{jk21}. We consider 54 Mica2Dot sensor nodes which are deployed in three dimensional space to collect temperature data in the network. In our simulations, 54 sensor nodes collects 800 observations done on 28th February 2004.

In the first simulation setup, our goal is to form clusters using 3D-DC algorithm in 3D space. We assume that source event occurs in 3D-space such that it propagates its signal spherically where the volume of spherical event sensing range is $V_e$. The sensor nodes which lies within this range $V_e$ are able to extract the source event observations in 3D space. Within $V_e$, sensor nodes form clusters using 3D-DC algorithm having dodecahedron topology. The circumscribed spherical sensing range of radius of a node is given by $r_n \leq  \left( \theta \hspace{0.04in} log \left(\frac{1}{\tau_n} \right) \right)^{1/\alpha}$  where the vertices's of dodecahedron touches the spherical sensing range such that $r_n=\frac{\nu \sqrt{3}}{4}(1+ \sqrt{5})\approx\left( \theta \hspace{0.04in} log \left(\frac{1}{\tau_n} \right) \right)^{1/\alpha}$ according to Algorithm 1. In our simulations, we consider $\tau_n=0.85$ where $0<\tau \leq 1$, $\theta=30$, $\alpha=1$ to calculate the radius of the dodecahedron clustered topology. The sensing coverage radius of each node is assumed to be 6 mts. We capture the 3D coordinates of each node in TABLE I. We assume all the sensor nodes falls within $V_e$ form a cluster. According to Algorithm 1, 3D-DC algorithm creates distributed clusters in 3D space. In TABLE II, seven dodecahedron clusters are formed using 3D-IE algorithm. Each dodecahedron cluster have a CH node with its associate neighboring sensor nodes. In each dodecahedron cluster, neighboring sensor nodes collaboratively transmits sensor data to its respective fusion node where fusion node estimates information accuracy for that cluster. From Table I, it is clear that the sensor nodes which are close to each other can form clusters as shown in TABLE II, based on their 3D coordinates. For example, in Table II, the cluster number III have fusion node 47 along-with associate nodes 5,6,12,28,32,38 who's 3D coordinates are almost close to each other illustrated in TABLE I, Therefore, they form cluster. Similarly, cluster number VI, consists of only a single fusion node id-16 who's coordinates are $x=9.612, y=9.591, z=9.373$. The coordinates of node id-16 is approximately far way from other nodes. Hence, no sensor nodes are associated with node id-16. In Table II, we compare our information accuracy done using 3D-IE algorithm with the existing models \cite{jk6}, \cite{jk7}, \cite{jk8}, \cite{jk11}. Result shows that 3D-IE algorithm does better estimation compared to existing models at the CH node of each dodecahedron cluster. In cluster number VI and VII, the fusion node id- 16 \& 11 shows information accuracy still they are not associated with other nodes due to the effect of sensing coverage from nearby clusters. Moreover, the table shows that the information accuracy achieved by cluster number VI is much lesser than other cluster as it has no associated nodes whereas cluster VII have somehow more information accuracy due to sensing effect of other clusters.  

\begin{table}[t!]
\centering
\caption{Sensor position (i.e x-y-z coordinates) in 3D space}
\label{Tabjk3} 

\begin{tabular}{|c|c|c|c|c|c|c|c|}
\hline
\textbf{Node id}  & \textbf{x}  & \textbf{y} & \textbf{z}  & \textbf{Node id}    & \textbf{x} & \textbf{y} & \textbf{z}        \\ \hline

1            &1.807   &6.525    &8.785        &28   &5.829    &9.513    &6.605      \\ \hline

2              &1.938    &9.937   &3.874      &29  &0.154    &0.462    &5.052         \\ \hline

3             &3.605    &7.137    &2.464          &30   &2.936   &0.209    &1.980            \\ \hline

4             &4.043    &1.023    &1.117        &31  &0.452    &0.880    &1.043             \\ \hline

5              &2.257    &4.472    &8.467         &32   &7.947    &5.178    &8.884           \\ \hline

6            &6.690    &6.927   &6.596         &33    &7.113    &8.786    &3.629         \\ \hline

7            &9.572    &3.329    &7.796         &34     &5.363    &4.021    &4.537       \\ \hline

8            &4.316    &9.217    &3.585        &35   &5.867    &5.613    &0.647         \\ \hline

9            &8.038    &6.324   &1.841      &36  &2.109    &5.981    &8.507      \\ \hline

10           &7.981    &9.321    &0.101          &37    &4.139    &0.750    &2.794            \\ \hline

11           &9.952    &4.259    &0.859       &38   &5.158   &3.914    &9.119         \\ \hline

12           &2.118    &3.004    &3.294        &39   &8.383    &3.549    &0.597           \\ \hline

13           &3.290    &8.890   &3.008        &40   &6.599    &2.447    &6.159         \\ \hline

14           &7.623    &0.173    &5.065        &41  &4.811    &1.080    &5.639            \\ \hline

15           &7.567     &1.480    &3.866       &42   &8.527    &1.122    &6.682          \\ \hline

16             &9.612    &9.591    &9.373      &43   &4.815    &2.672    &3.354           \\ \hline

17             &5.698    &7.144    &1.595         &44    &2.723    &2.663    &0.377            \\ \hline

18            &6.085    &3.064    &7.797      &45    &7.312    &9.360    &2.857            \\ \hline

19             &3.650    &8.281   &1.774        &46   &9.694   &1.861    &9.312          \\ \hline

20             &3.435    &8.078    &8.052       &47   &3.756    &5.074    &7.998          \\ \hline

21             &5.999    &9.092    &5.623       &48    &2.386    &1.475    &3.310           \\ \hline

22           &5.256    &6.428    &4.700         &49    &2.427    &9.207    &7.374            \\ \hline

23            &4.849    &6.286    &3.596       &50   &0.189    &9.294    &5.674           \\ \hline

24              &6.556    &1.184    &3.717       &51   &9.826    &1.367    &0.160           \\ \hline

25             &0.382    &9.189    &3.828       &52   &8.026    &8.715    &4.255       \\ \hline

26             &1.899    &6.239    &2.563      &53  &5.634    &0.123    &0.293        \\ \hline

27              &9.287    &2.575    &9.720     &54   &3.888    &7.220    &7.841           \\ \hline

\end{tabular}
\end{table}

\begin{table*}[ht]
  \begin{tabular}{*{17}{cc}}

 \hline\hline

Cluster  &  Fusion     &  Associated nodes           & Accuracy  & Accuracy   & Accuracy   & Accuracy        &Accuracy  \\   [1ex]                                                                              Number   &  Node       &   ID in cluster             &Vuran at. all \cite{jk6}  & Li at. all \cite{jk8}  & Chai at. all\cite{jk7} & Karjee at. all\cite{jk11} & 3D-IE   \\  [0.2ex]

\hline \\
       I    & 25      & 1,2,3,8,13,19,20,21            &0.9412    &0.9412    &0.9300    &0.9310    &0.9424   \\
        
                 &       &22,23,26,36,49,50,54                              &             &                 &     &         &           \\
  \hline \\

       II & 14     &4,7,15,18,24,27,30,34,37,             & 0.9570    &0.9570    &0.9484    &0.9490    &0.9593 \\

                &       &39,40,41,42,43,46,48,51,53                          &           &                &   &          &               \\

  \hline \\

     III & 47     &5,6,12,28,32,38                &0.9656    &0.9656    &0.9566    &0.9582    &0.9676  \\

\hline \\

     IV  & 33     &9,10,17,35,45,52                          &0.9678    &0.9678    &0.9531    &0.9556    &0.9730 \\

\hline \\

     V & 31    &29,44                             &0.9604    &0.9604    &0.9495    &0.9539    &0.9641 \\

  \hline \\

     VI & 16    & -                                  &0.9176    &0.9176    &0.9121    &0.9176    &0.9244 \\

   \hline \\

     VII & 11    & -                        &0.9427    &0.9427    &0.9376    &0.9427    &0.9460 \\

   \hline \\

 \end{tabular}

\caption{Clustering using 3D-DC with 3D-IE compared with existing models}
\end{table*}

In the second simulation setup, we are interested to find the node placement strategy of sensor nodes using 3D-NP algorithm. We assume that all the 54 sensor nodes are with in the sensing range $V_e$. To achieve this goal, we need to perform the experiments under spatio-temporal effects of information accuracy. To validate the information correlation among sensor nodes in 3D space, we simulate to identify the sensor nodes which can maximize information accuracy in 3D space. This means, the sensor nodes that have more signal variance and covariance can be selected in the network. A cost function is calculated for node deployment scenario which perform the information accuracy level for the dedicated 3D networks. Moreover, we validate the temporal information correlation among sensor nodes in 3D space. 3D-NP algorithm searches for the nodes which maximizes information accuracy w.r.t to time using the cost function. As the time progresses, the cost function have temporal effect on the 3D network, as it reaches a saturation level after certain duration. In Figure 4, the sensor nodes having higher cost function are being selected from 54 nodes, which actually maximize the information accuracy in the network. The nodes with higher cost function have more signal variances and covariances, where as nodes having lower cost function have less variances and covariances. For example, sensor nodes deployed directly under sunlight have higher cost function w.r.t to nodes deployed under shade of trees, still they are placed under same geographical region. Here in the tree dominated sensing region, the cost function is low. To maximize information accuracy in such tree dominated sensing region, it is recommended to put more nodes to maximize the information accuracy. Therefore, the sensor node ids: 1 to 23 and node ids: 33 to 47 are selected, considering a benchmark of minimum cost function threshold value of 5. Above this benchmark, nodes have higher variances and covariances values and are selected for maximizing information accuracy for the network. Therefore in Figure 4, out of 54 sensor nodes, 37 sensor nodes are selected to maximize information accuracy. In Figure 5, the cost function achieves a desired information accuracy level as the number of sensor nodes increases. Cost function shows that approximately 38 sensor nodes are sufficient to give a desirable information accuracy as validated by Figure 4.  

In Figure 6, cost function searches for the nodes having maximum information accuracy with respect to time (or number of rounds). We perform 300 rounds to select the nodes (as given in Figure 4), which have higher variances or covariances of signal above a desired benchmark threshold value. Similarly, in Figure 7, the cost function achieves a saturation level as the time progresses. This means, transmitting more number of data packet to the sink node doesn't give extra advantage over cost function. Since, it has temporal effect on 3D networks, transmitting a subset of information is sufficient to achieve the desired information accuracy as number of round increases.

\begin{figure}
\centering
\includegraphics[width=0.47\textwidth]{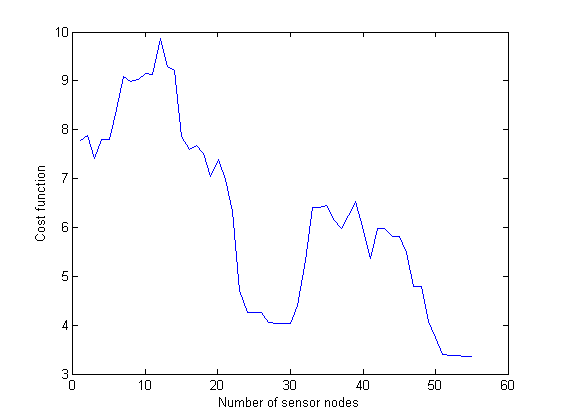}
\caption{Cost function versus number of sensor nodes}
\end{figure}

\begin{figure}
\centering
\includegraphics[width=0.47\textwidth]{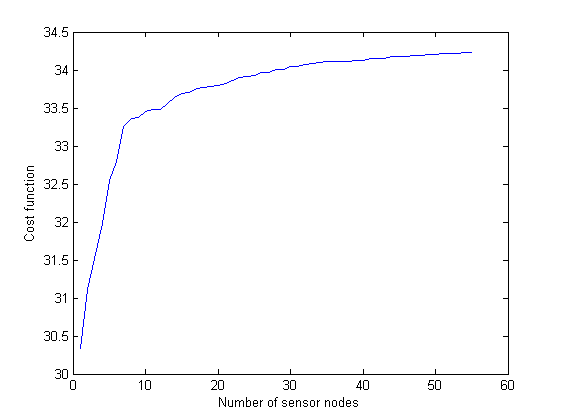}
\caption{Cost function versus number of sensor nodes}
\end{figure}

\begin{figure}
\centering
\includegraphics[width=0.5\textwidth]{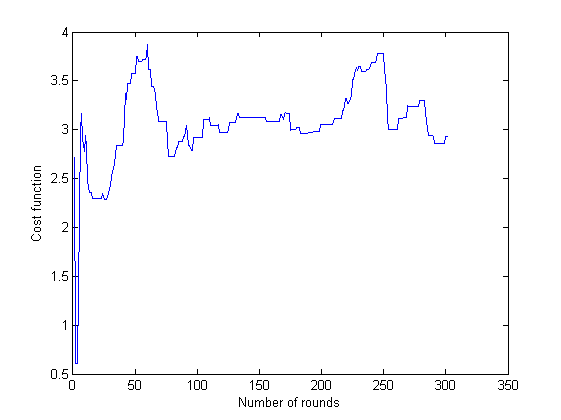}
\caption{Cost function versus number of rounds}
\end{figure}

\begin{figure}
\centering
\includegraphics[width=0.5\textwidth]{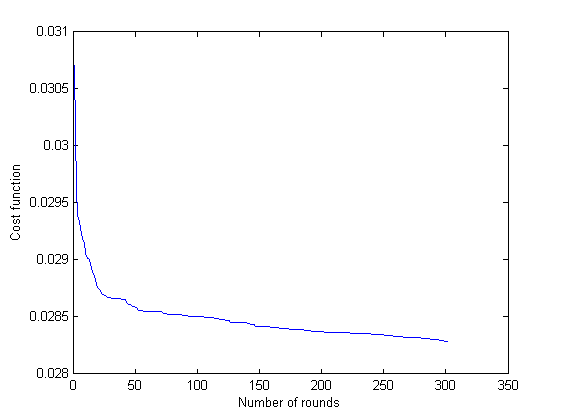}
\caption{Cost function versus number of rounds}
\end{figure}

\section{Conclusions}

A sensing model for 3D-WSN based dodecahedron topology called 3D-DC algorithm is developed to form cluster, where there is no tessellate space among them. In each dodecahedron cluster, Cluster Head (CH) node extracts accurate estimates of information using Three Dimensional Information Estimation (3D-IE) algorithm. Moreover, node deployment is an important factor to maximize information accuracy in 3D-WSN. We consider node deployment scenario using 3D-NP algorithm which can find an optimal way of placing sensor nodes in a sensing field to extract maximum information accuracy in the network. We validate 3D-NP algorithm using simulation results. 3D-DC, 3D-IE and 3D-NP algorithms may reduce communication overhead and energy consumption in sensor networks.

\section{Acknowledgement}
This work was carried out by the authors at Indian Institute of Science, India. The work was submitted, while the corresponding author was associated with Manipal University Jaipur, India. The revised version of this paper is submitted while the corresponding author is associated with the above present address.

\begin{IEEEbiography}[{\includegraphics[width=1in,height=1.25in,clip,keepaspectratio]{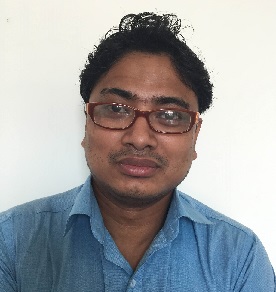}}]{Jyotirmoy Karjee}
received his PhD in Engineering from Indian Institute of
Science, Bangalore, India. He did his Post-Doctoral research from Technische Universität München, Germany. Presently, he holds the position of Researcher at Embedded Systems and Robotics group, TCS Research and Innovation, Bangalore, India. He is a recipient of Heritage Erasmus Mundus scholarship (fellowship) to do postdoctoral research. He is interested in statistical signal processing, wireless sensor networks, wireless communications, robotic communications and embedded systems, machine learning, cloud computing and Internet of things.
\end{IEEEbiography}

\begin{IEEEbiography}[{\includegraphics[width=1in,height=1.25in,clip,keepaspectratio]{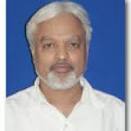}}]{H.S Jamadagni}
received his M.E and Ph.D degree in Electrical and Communication Engineering from Indian Institute of Science, Bangalore. Currently, He is the professor at Department of Electronic System Engineering, Indian Institute of Science. He is one of the main coordinators for the Intel higher education program, member of Telecommunications Regulatory Authority of India (TRAI) and key mentors for various intel workshops in india. His current research work includes in the areas of communication networks, embedded systems, VLSI for wireless networks, etc.
\end{IEEEbiography}


\begin{thebibliography}{1}

\bibitem{jk1}
I.F Akyuildz, W. Su,Y. Sankarasubramanian, E.Cayirci ,``A Survey on sensor Networks", \emph {IEEE Communica
tions Magazine} ,vol.40, pp.102-104, Aug 2002.


 \bibitem{jk2}
J. Kahn, R. Katz and K. Pister,``Next Century Challenges: Mobile Networking for Smart Dust", \emph {Proceedings of 5th Annual International Conf. on Mobile Computing and Networking (Mobicom)}, pp.263-270, 1999.

\bibitem{jk3}
B. Warneke, M. Last, B. Leibowitz and K. Pister, ``Smart Dust: Communicating with cubic-millimeter computer", \emph {IEEE computer Magazine} vol.34, pp.44-51, January 2001.

\bibitem{jk4} I.F Akyildiz, D. Pompili and T. Melodia, ``Underwater Acoustic Sensor Networks: Research Challenges", \emph {Elsevier Jounrnal on Adhoc Networks}, vol 3, pp. 257-279,2 005.


\bibitem{jk5}
Steere D, Baptista A, Mcnamee D, Pu C and Walpole J,``Research Challenges in Environmental Observation and Forecasting Systems", \emph {Proceedings of the 6th annual international conference on Mobile computing and networking (MobiCom)}, pp. 292-299, Boston, USA, 2000.


\bibitem{jk6}
  M.C Vuran, O.B Akan, I.F Akyildiz,
  ``Spatio-Temporal Correlation: Theory and Applications for Wireless Sensor Networks", \emph{Computer Networks, Elsevier}, pp. 245-259, 2004.


  \bibitem{jk7}
  K.Cai,G. Wei, H. Li, ``Information accuracy versus jointly sensing nodes in wireless sensor networks ",
  \emph{IEEE Asia Pacific Conference on Circuits and Systems }, pp. 1050-1053, 2008.


  \bibitem{jk8}
  H.Li, S. Jiang, G. Wei,
 ``Information accuracy aware jointly sensing nodes selection in wireless sensor networks",
  \emph{LNCS, Springer }, pp. 736-747, Verlag Berlin, 2006.


 \bibitem{jk9}
Jyotirmoy karjee , H.S jamadagni ,``Optimal Node Selection using Estimated Data Accuracy Model in
Wireless Sensor Networks", \emph {Third International conference on Recent Trends in Information, Telecommunication and Computing, Lecture Notes in Electrical Engineering}  Vol. 150, pp 195-205, 2012.


\bibitem{jk10}
Jyotirmoy karjee, H.S Jamadagni,``Energy Aware Node Selection for Cluster-based Data accuracy Estimation in wireless sensor networks" \emph {International Journal of Advanced Networking and Applications}, vol-3, Issue 5, pp. 1311-1322, 2012.


\bibitem{jk11}
Jyotirmoy Karjee, H.S Jamadagni ,``Data accuracy Estimation for Spatially Correlated Data in Wireless
Sensor networks under Distributed Clustering", \emph {Journal of Networks}, vol-6 , no.7, pp 1072-1083, 2011.


\bibitem{jk12}
Chongqing Zhang, Binguo Wang, Sheng Fang , Jiye Zheng, ``Spatial Data Correlation Based Clustering Algorithms for Wireless Sensor Networks", \emph{The 3rd International Conference on Innovative Computing Information and Control}, 2008.


\bibitem{jk13}
Zhikui chen, Song Yang , Liang Li and Zhijiang Xie ,
``A clustering Approximation Mechanism based on Data Spatial Correlation in Wireless Sensor Networks", \emph{Proceedings of the 9th International Conference on Wireless Telecommunication Symposium}, 2010.


\bibitem{jk14}
Jyotirmoy Karjee, H.S Jamadagni, ``Data Accuracy Models under Spatio-Temporal Correlation with Adaptive Strategies in wireless Sensor Networks", \emph{ACEEE International Journal on Network Security}, Vol.4, No.1, July 2013.


\bibitem{jk1411}
Jyotirmoy Karjee, Martin Kleinsteuber, ``Data estimation with predictive switching mechanism in wireless sensor networks", International Journal of Sensor Networks, \emph{paper in press}.


\bibitem{jk15}
Jyotirmoy Karjee, Martin Kleinsteuber, H.S Jamadagni, ``Spatial Data Estimation in Three Dimensional Distributed Wireless Sensor Networks", \emph{International Conference on Embedded Systems}, pp-139-144, 2014.

\bibitem{jk16}
Manas Kumar Mishra, M. M. Gore,
``On Optimal Space Tessellation with Deterministic Deployment for Coverage in Three-Dimensional Wireless Sensor Networks", \emph{ Lecture Notes in Computer Science}, Vol. 5966, pp 72-83, 2010.

\bibitem{jk17}
Fu Xiao, Yang Yang, Ruchuan Wang and Lijuan Sun,
``A Novel Deployment Scheme Based on Three-Dimensional Coverage Model for Wireless Sensor Networks",
\emph{The Scientific World Journal}, Vol. 2014, Article ID 846784, 7 pages, 2014.

\bibitem{jk18}
J.O Berger, V.de Oliviera and B. Sanso , ``Objective Bayesian Analysis of Spatially Correlated Data", \emph {Journal of American Statistical Association}, Vol-96, pp.1361-1374, 2001.

\bibitem{jk19}
De Olivera V, Kedan B and Short D.A , ``Bayesian prediction of transformed Gaussian Random Fields ",\emph {Journal of American Statistical Association}, vol-92, pp.1422-1433, 1997.


\bibitem{jk20}
A.H Syed, ``Adaptive Filters",\emph{ John Wiley and Sons,NJ}, 2008.

\bibitem{jk21}
Intel Lab Data: http://db.csail.mit.edu/labdata/labdata.html

\bibitem{jk22}
Michael Sherman, ``Spatial Statistics and Spatio-Temporal Data: Covariance Functions and Directional Properties", \emph{ Wiley}, 2010.

\bibitem{jk23}
X. Hu, R. Eberhart and Y. Shi, ``Recent advances in particle Swarm", \emph{  IEEE congress on Evolutionary Computation}, pp. 90-97, 2004. 

\bibitem{jk24}
Kennedy, J. and Eberhart, R. C., ``Particle swarm optimization", \emph{ Proceedings of IEEE international conference on neural networks},  Vol. IV, pp. 1942-1948, IEEE service center, Piscataway, NJ, 1995.

\bibitem{jk25}
Eberhart, R. C. and Kennedy, J., ``A new optimizer using particle swarm theory", \emph{ Proceedings of the sixth international symposium on micro machine and human science} pp. 39-43. IEEE service center, Piscataway, NJ, Nagoya, Japan, 1995.

\bibitem{jk26}
Eberhart, R. C. and Shi, Y. ``Particle swarm optimization: developments, applications and resources", \emph{Proc. congress on evolutionary computation 2001 IEEE service center}, Piscataway, NJ., Seoul, Korea., 2001.

\bibitem{jkA27}
Bang Wang, Jiajun Zhu, Laurence T. Yang, Yijun Mo, ``Sensor Density for Confident Information Coverage in Randomly Deployed Sensor Networks", \emph{ IEEE Transactions on Wireless Communications}, vol.15, no.5, pp.3238-3250, 2016.

\bibitem{jkA28}
Jiajun Zhu, Bang Wang, ``The Optimal Placement Pattern for Confident Information Coverage in Wireless Sensor Networks", \emph{IEEE Transactions on Mobile Computing}, vol. 15, no.4, pp.1022-1032, 2016.

\bibitem{jkA29}
Bang Wang, Han Xu, Wenyu Liu, Laurence T. Yang, ``The Optimal Node Placement for Long Belt Coverage in Wireless Networks",\emph{ IEEE Transactions on Computers}, vol. , no. , pp.587-592, 2015.


\bibitem{jkA30}
 Bang Wang, Xianjun Deng, Wenyu Liu, Laurence T. Yang, Han-Chieh Chao  ``Confident Information Coverage in Sensor Networks for Field Reconstruction", \emph{IEEE Wireless Communications}, vol.20, no.6, pp.74-81, 2013.
   

\bibitem{jk301}
Jyotirmoy Karjee, H.S Jamadagni, ``Data Accuracy Model for Distributed Clustering Algorithm based on Spatial Data Correlation in Wireless Sensor Networks”, \emph{ GESJ:Computer Sciences and Telecommunications}, No-2(34), pp.13-27, 2012.

\bibitem{jk302}
Jyotirmoy Karjee, H.S Jamadagni, ``Data accuracy estimation for cluster with spatially correlated data in wireless sensor networks”, \emph{IEEE International Conference on Communication Systems and Network Technologies}, pp-427-434, 2012.


\bibitem{jkA31}
Bang Wang, ``Coverage Problems in Sensor Networks: A Survey", \emph{ACM Computing Surveys}, vol.43, no.4, pp.1-56, Oct., 2011.

\bibitem{jkA32}

Bang Wang, Kee Chaing Chua, Vikram Srinivasan, ``Information Coverage in Randomly Deployed Wireless Sensor Networks", \emph{IEEE Transactions on Wireless Communications}, vol.6, no.8, pp.2994-3004, 2007.



\end{thebibliography}
\end{document}